\documentstyle[11pt]{article}
\begin{document}
\begin{titlepage}
\begin{centering}
 
{\ }\vspace{2cm}
 
{\Large\bf Annular Vortex Solutions to the Landau-Ginzburg Equations}\\
%\vspace{0.5cm}
%{\Large\bf in}\\
\vspace{0.5cm}
{\Large\bf in Mesoscopic Superconductors}\\
\vspace{2cm}
Jan Govaerts, Geoffrey Stenuit, Damien Bertrand and Olivier van 
der Aa\footnote{E-mail addresses:
{\tt govaerts, stenuit, bertrand, vanderaa @fynu.ucl.ac.be}}\\
\vspace{1.0cm}
{\em Institut de Physique Nucl\'eaire}\\
{\em Universit\'e catholique de Louvain}\\
{\em 2, Chemin du Cyclotron}\\
{\em B-1348 Louvain-la-Neuve, Belgium}\\
\vspace{2cm}
\begin{abstract}

\noindent New vortex solutions to the Landau-Ginzburg equations are described.
These configurations, which extend the well known Abrikosov and giant magnetic
vortex ones, consist of a succession of ring-like supercurrent vortices 
organised in a concentric pattern, possibly bound to a giant magnetic vortex 
then lying at their center. The dynamical and thermodynamic stability of these 
annular vortices is an important open issue on which hinges the direct 
experimental observation of such configurations. Ne\-ver\-the\-less, annular 
vortices should affect indirectly specific dynamic properties of mesoscopic 
superconducting devices amenable to physical observation.

\vspace{60pt}

\end{abstract}

\end{centering} 

\vspace{80pt}
%\vspace{30pt}


\vspace{20pt}

%\noindent cond-mat/9908xxx\\
\noindent August 1999

\end{titlepage}

\setcounter{footnote}{0}

\noindent{\bf 1. Introduction.}
Ever since they were introduced, the Landau-Ginzburg (LG) 
equations\cite{Tin} have remained a powerful and insightful tool for 
the exploration and understanding of superconducting and other collective 
quantum phenomena, in particular in low temperature superconductors. 
Among a host of important results and developments which were
achieved through the use of these equations, the existence of Abrikosov 
magnetic vortex solutions\cite{Tin} still stands out as a most remarkable 
prediction transcending the boundaries of condensed matter physics
by itself, whose physical significance and applications still have
many riches in store even forty years on, especially with the recent
advent of mesoscopic and nanoscopic superconducting devices.

In this letter, we wish to report on new solutions to the LG equations
which share some properties with usual Abrikosov configurations,
but are nevertheless of quite a different nature and are possible because
of the boundary conditions implied by mesoscopic samples
of finite size. Rather than having the order parameter vanish along a line 
within the superconduc\-ting material as occurs for Abrikosov and giant 
magnetic vortices, these solutions are characterized by having the order 
parameter vanish on one or more {\em surfaces\/} each with the topology of 
a cylinder and enclosing one another in a concentric-like pattern with more 
or less constant spacing, while the maximal number of 
such surfaces is set by the size of the mesoscopic sample relative to the
coherence length $\xi $ of the material\footnote{A few weeks after these 
solutions had been found, we received Ref.\cite{Tris} in which similar
ones are discussed in the context of quantum optics, based on
coupled non linear equations analogous to the LG ones.}. Moreover,
the volume defined by any pair of successive surfaces is threaded by
a supercurrent running in a closed loop,
parallel to and vanishing on these surfaces, with
all these successive sheaths of currents having the same circular orientation.
Finally, these configurations may also carry a non zero fluxoid quantum
number, in which case they may be viewed as being bound to
an ordinary giant vortex carrying that quantum of flux and lying at the
center of the set of surfaces, the associated linear locus of the vanishing
order parameter then threading along the central axis of the innermost surface.
Because of this particular topology, we refer to these solutions as
``annular vortices"\footnote{The idea of a vortex being associated to that of
a current running in a closed loop.} of order $n$ and fluxoid $L$,
$n$ referring to the number of cylindrical surfaces on which the order
parameter vanishes and $L$ being the usual fluxoid quantum number. These 
solutions thus generalise the well known giant magnetic vortex ones
which correspond to $n=0$ and an arbitrary number $L$ of Abrikosov vortices
or anti-vortices lying on top of one another\cite{Vega}.

Since these configurations solve the LG equations, they define local
extrema of the free energy, but it has not yet been established whether
these are local minima rather than maxima or saddle points
whatever the values for $n$ and $L$. 
Irrespective of this issue of their dynamical stability in 
configuration space\footnote{Note that the stability of the ordinary 
giant vortex solutions $(L,n=0)$ seems to have been settled
only recently\cite{Gustafson}.}, it is clear that for a fixed value of $L$,
the free energy of these solutions increases with increasing $n$.
Hence in practical physics terms, such configurations have only a finite
thermodynamic lifetime which also still needs to be assessed.

Consequently, at this stage, it is not clear whether these annular
vortices could be observed in mesoscopic devices, and how low
a temperature would be required for that purpose. Nevertheless, from
the quantum mechanical point view, annular vortices---even if dynamically
unstable---provide for new contributions to the density of states
of such systems, and should therefore affect specific physical
properties of mesoscopic devices. One such instance could well be the
behaviour in an external magnetic field of mesoscopic disks and 
annuli\cite{Geim1}, whose maze of $(L,n=0)$ energy curves in their
phase diagram is then filled by many more such curves for $n\ne 0$,
leading to a plethora of degeneracy crossings as a function of the
applied field and thereby a dynamics of these devices possibly much richer 
still than that so far considered\cite{Bruyn1,Bruyn2}.

In fact, having in mind eventual applications such as a new type
of quantum detector for elementary particles and atoms,
or as the basic qubit device for a quantum computer, these annular
vortex solutions were found precisely when studying mesoscopic samples
with a cylindrical and annular geometry, and assuming an infinite
extent along their symmetry axis. In that case, in a plane
perpendicular to that axis, the order parameter vanishes at the center
when $L\ne 0$, while a series of almost regularly spaced $n$ concentric circles
of vanishing order parameter appear when $n\ne 0$ (in the case of an annulus,
the same description applies with the central hole simply removed).
This is the situation presented in this letter, but it is clear that
for samples whose boundaries are of arbitrary shape, the same classes
of solutions exist, obtained through continuous deformations of
the perfectly symmetric ones constructed here, thereby leading to the
general picture for these annular vortices as described above.

\noindent{\bf 2. The LG equations.}
For time independent configurations in the presence of an external magnetic 
field $\vec{B}_{\rm ext}$, in S.I. units the free energy of the system 
writes as\cite{Gov1}
\begin{equation}
\begin{array}{r c l}
E&=&\int_{(\infty)}d^3\vec{x}\frac{1}{2\mu_0}
\left(\vec{B}-\vec{B}_{\rm ext}\right)^2\ +\\
&+&\int_\Omega d^3\vec{x}\frac{1}{2\mu_0}
\left(\frac{\Phi_0}{2\pi\lambda^2}\right)^2\,\lambda^2\,
\left\{\left|\vec{\nabla}\psi-i\frac{q}{\hbar}\vec{A}\psi\right|^2+
\frac{1}{2\xi^2}\left[|\psi|^2-1\right]^2-\frac{1}{2\xi^2}\right\}\ \ \ .
\end{array}
\label{eq:FreeE}
\end{equation}
The notation used should be self-explanatory for most variables.
The symbol $\Omega$ stands for the volume of the superconducting sample,
$\Phi_0=2\pi\hbar/|q|$---with $q=-2e$ ($e$ being the positron charge)---is
the usual unit of quantum of flux, while $\lambda$ and $\xi$ are
the penetration and coherence lengths, respectively.
However, we need to emphasize the fact that in this expression
the order parameter $\psi(\vec{x})$ is normalised to its constant value
in the bulk of the material in the absence of any magnetic field, and that
the free energy is defined such as to vanish exactly at the 
normal-superconducting transition.

The above expression of the free energy makes it clear that
it is the penetration length $\lambda$ which weighs the relative contributions
of the magnetic field energy---due to deviations of $\vec{B}(\vec{x})$ from the
applied field $\vec{B}_{\rm ext}$---and of the condensate energy---due
to deviations of $|\psi(\vec{x})|$ from zero---, while it is the coherence 
length $\xi$ which weighs the contributions to the condensate energy due
to spatial inhomogeneities in $\psi(\vec{x})$---through the covariantized
gradient---relative to those due to deviations from the bulk value
$|\psi|=1$---through the LG potential energy. This interplay of scales
is made manifest in (\ref{eq:FreeE}).

Extremizing the free energy leads to the LG equation
\begin{equation}
\lambda^2\left[\vec{\nabla}-i\frac{q}{\hbar}\vec{A}\right]^2\psi=
\kappa^2\,\psi\left[|\psi|^2-1\right]\ \ \ ,
\end{equation}
$\kappa=\lambda/\xi$ being the LG parameter, coupled the Maxwell equation
\begin{equation}
\vec{\nabla}\times\vec{B}=\mu_0\,\vec{J}_{\rm em}\ \ \ ,
\end{equation}
with the electromagnetic current density given by
\begin{equation}
\mu_0\vec{J}_{\rm em}=\frac{i}{2}\frac{\Phi_0}{2\pi\lambda^2}\,
\Bigg[\psi^*\left(\vec{\nabla}\psi-i\frac{q}{\hbar}\vec{A}\psi\right)-
\left(\vec{\nabla}\psi^*+i\frac{q}{\hbar}\vec{A}\psi^*\right)\psi\,\Bigg]\ \ \ ,
\ \ \ \vec{\nabla}\cdot\left(\mu_0\vec{J}_{\rm em}\right)=0\ \ \ .
\label{eq:Jem}
\end{equation}
In particular, this latter relation implies the second London equation
for the Meissner effect
\begin{equation}
\vec{\nabla}\times\left(\frac{\lambda^2}{|\psi|^2}\mu_0\vec{J}_{\rm em}\right)=
-\vec{B}\ \ \ .
\end{equation}

Let us now particularize these equations to axially symmetric configurations, 
the external homogeneous field $\vec{B}_{\rm ext}$ being also aligned with
the symmetry axis $\hat{e}_z$. We shall specifically consider a 
sample either of cylindrical geometry with radius $r_b$ (and $r_a=0$)
or of annular geometry with inner and outer radii $r_a$ and $r_b$, respectively. 
Finally, in the present study we assume the extension of the sample in the 
$\pm\hat{e}_z$ directions to be infinite. The following parametrisations 
then apply,
\begin{equation}
\vec{B}=B(r)\hat{e}_z\ ,\ 
\vec{A}=A(r)\hat{e}_\phi\ ,\ 
\mu_0\vec{J}_{\rm em}=J(r)\hat{e}_\phi\ ,\ 
\psi(r,\phi)=f(r)e^{i\theta(\phi)}\ ,\ 
\rho(r)=|\psi(r,\phi)|^2=f^2(r)\ ,
\end{equation}
where $(r,\phi)$ define the polar coordinates in the plane
perpendicular to the $\hat{e}_z$ symmetry axis. In contradistinction
to other works, an important point needs to be emphasized concerning the
parametrisation used for the order parameter in terms of the two functions
$f(r)$ and $\theta(\phi)$. The function $f(r)$ is of course real,
{\em but of a sign which may be positive or negative as the case
arises\/}, while the phase function $\theta(\phi)$ is assumed to
be {\em continuous\/}. Usually, the function $f(r)$ is explicitely assumed
to be the square-root of the (normalised) condensate density $\rho(r)$,
$f(r)=\sqrt{\rho(r)}$, whose sign is thus always positive,
with the important consequence that whenever the order parameter vanishes 
such that the gradiant of $\sqrt{\rho}$ does not vanish as well, 
on the ``other side of the zero" there is necessarily a discontinuous jump 
of $\pm\pi$ in the phase factor $\varphi$ which is then used to parametrise 
the order parameter through $\psi=\sqrt{\rho}e^{i\varphi}$. 
There is of course no ``other side of the
zero" in the case of a giant vortex, but there is obviously one when
$\psi$ vanishes on a surface within the sample\footnote{From that point of view,
annular vortices are very much reminiscent of domain wall configurations in the
spontaneously broken Yang-Mills theories of particle physics and cosmology,
and could thus possibly be of interest in these other fields 
of physics as well, not to mention of course other coherent quantum systems in
condensed matter physics.}. The general parametrisation of $\psi(\vec{x})$ 
advocated here allows for the specific possibility that $f(\vec{x})$ may 
change sign when crossing a zero of 
$\psi(\vec{x})=f(\vec{x})e^{i\theta(\vec{x})}$, 
all in a continuous fashion for both functions $f(\vec{x})$ and 
$\theta(\vec{x})$ and their gradients. 
This point is obviously essential to our results.

To proceed further, let us choose to normalise quantities in terms of
the penetration length and the associated unit of magnetic field
$\Phi_0/(2\pi\lambda^2)$, by introducing the dimensionless variables,
\begin{equation}
u=\frac{r}{\lambda}\ \ ,\ \ 
b(u)=\frac{B(u)}{\Phi_0/(2\pi\lambda^2)}\ \ ,\ \ 
a(u)=\frac{1}{\lambda}\,\frac{A(u)}{\Phi_0/(2\pi\lambda^2)}\ \ ,\ \ 
g(u)=u\,\frac{q\lambda^3}{\hbar}\,\frac{1}{f^2(u)}\,J(u)\ \ .
\end{equation}
The coupled LG-Maxwell equations then reduce to\cite{Gov1}
\begin{equation}
\frac{1}{u}\frac{d}{du}\left[u\frac{d}{du}f(u)\right]=
\frac{1}{u^2}f(u)g^2(u)-\kappa^2f(u)\left[1-f^2(u)\right]\ \ ,\ \ 
u\frac{d}{du}\left[\frac{1}{u}\frac{d}{du}g(u)\right]=f^2(u)g(u)\ \ ,
\label{eq:LG2}
\end{equation}
while the magnetic field $B(r)$ is determined by the second London equation
\begin{equation}
b(u)=\frac{1}{u}\frac{d}{du}g(u)\ \ \ .
\end{equation}
Furthermore, a careful consideration of the boundary conditions (b.c.) of the
problem, of the small $u$ behaviour of solutions to these equations, of the
flux threading the sample and of the relation (\ref{eq:Jem}) between
$\mu_0\vec{J}_{\rm em}$ and $\psi$, shows that the remaining quantities
are determined to be\cite{Gov1}
\begin{equation}
\theta(\phi)=-L\phi+\theta_0\ \ \ ,\ \ \ ua(u)=g(u)+L\ \ \ ,\ \ \
L=0,\pm 1,\pm 2,\dots\ \ \ ,
\end{equation}
where $\theta_0$ is arbitrary and $L$ is the usual fluxoid quantum number,
thus corresponding to the order parameter 
$\psi(r,\phi)=f(r)e^{-iL\phi}e^{i\theta_0}$ characteristic of giant
vortex configurations for $L\ne 0$.
In particular, the total flux threading a disk of radius $u_a<u<u_b$ 
within the sample is
\begin{equation}
\Phi(u)=\Phi_0\,\left[g(u)+L\right]=\Phi_0\,ua(u)\ \ \ ,\ \ \ 
u_a<u<u_b\ \ \ .
\end{equation}
The advantage of expressing the problem in these terms rather than
those usual in the literature, is that only
{\em gauge invariant\/} quantities still appear in the coupled equations
to be solved, while the gauge variant ones, $\theta(\phi)$ and
$ua(u)$, are explicitely known already.

The relevant b.c. still need to be specified.
In the disk case surrounded by the vacuum or an insulating
material, they are\cite{Bruyn1,Gov1}
\begin{equation}
\begin{array}{l c l c l}
{\rm at}\ u=0&:& g(u)_{|_{u=0}}=-L &;& 
\frac{df(u)}{du}_{|_{u=0}}=0\ {\rm if}\ L=0\ \ {\rm or}\ \ 
f(u)_{|_{u=0}}=0\ {\rm if}\ L\ne 0\ \ ,\\
{\rm at}\ u=u_b&:& \frac{1}{u}\frac{dg(u)}{du}_{|_{u=u_b}}=b_{\rm ext} &;&
\frac{df(u)}{du}_{|_{u=u_b}}=0\ \ ,
\end{array}
\label{eq:bcdisk}
\end{equation}
while in the annulus case, we have\cite{Gov1}
\begin{equation}
\begin{array}{l c l c l}
{\rm at}\ u=u_a&:& \frac{1}{2}u_a\frac{dg(u)}{du}_{|_{u=u_a}}=g(u_a)+L &;&
\frac{df(u)}{du}_{|_{u=u_a}}=0\ \ ,\\
{\rm at}\ u=u_b&:& \frac{1}{u}\frac{dg(u)}{du}_{|_{u=u_b}}=b_{\rm ext} &;&
\frac{df(u)}{du}_{|_{u=u_b}}=0\ \ .
\end{array}
\label{eq:bcann}
\end{equation}

Finally, it may be worth noting also that both these sets of b.c.
as well as the coupled equations (\ref{eq:LG2}) for $f(u)$ and $g(u)$
derive from the variation of the following quantity (valid both for the
disk with $u_a=0$ and for the annulus with $u_a\ne 0$)
\begin{equation}
\begin{array}{r c l}
&&\frac{1}{2}\int_{u_a}^{u_b}du\,u\,
\left\{\left[\frac{df}{du}\right]^2+
\left[\frac{1}{u}\frac{dg}{du}-b_{\rm ext}\right]^2+
\frac{1}{u^2}f^2g^2+\frac{1}{2}\kappa^2\left(1-f^2\right)^2-\frac{1}{2}\kappa^2
\right\}\,+\,\\
&+&\frac{1}{4}\,u^2_a\,\left[2\frac{g(u_a)+L}{u^2_a}-b_{\rm ext}\right]^2\ =
\ \frac{1}{2}\,\frac{2\mu_0}{2\pi\lambda^2}\,
\left(\frac{2\pi\lambda^2}{\Phi_0}\right)^2\,{\cal E}\equiv{\cal E}_{\rm norm}
\ \ \ ,
\end{array}
\label{eq:FreeE2}
\end{equation}
where ${\cal E}$ denotes the free energy $E$
per unit length in the $\hat{e}_z$ direction of the sample.

\noindent{\bf 3. Analytical insight.}
The resolution of the equations (\ref{eq:LG2}) requires a numerical approach,
both because of the non linear character of these equations as well as
the b.c. at finite values of $r$. Nevertheless, some
analytical considerations may be developed, giving already
the insight necessary to the construction of annular vortices
and the actual demonstration of their existence.

The pair of coupled second order equations (\ref{eq:LG2}) requires a total
of four b.c.. However, in order to determine a solution
uniquely, these four conditions must be specified {\em at the same boundary\/},
thereby enabling the construction of the then unique solution by propagating
these boundary values throughout the sample volume using the equations
(\ref{eq:LG2}). In the present case, we do have a total of four such
conditions, but split evenly between the two boundaries of the sample
at $u_a$ ($u_a=0$ in the disk case) and $u_b$. In order to construct
solutions, one needs to add two free b.c. at either boundary,
propagate the solution within the sample and then adjust the two free
conditions in order to meet the two conditions specified at the other
boundary. Consequently, this procedure may generate more than a single
solution.

To be specific, consider first the disk case and introduce the
parametrisation
\begin{equation}
f(u)=u^{|L|}\tilde{f}(u)\ \ ;\ \
f_0=\tilde{f}(u)_{|_{u=0}}\ \ ,\ \ 
g_0=\frac{1}{2u}\frac{dg(u)}{du}_{|_{u=0}}=\frac{1}{2}\,b(u)_{|_{u=0}}\ \ .
\end{equation}
{}From the equations (\ref{eq:LG2}), it follows that $\tilde{f}(u)$ and
$g(u)$ are both even functions of $u$, while the two boundary
conditions at $u=0$ in (\ref{eq:bcdisk}) are then also satisfied. Thus,
the parameters $f_0$ and $g_0$ correspond to the two additional free
boundary values at $u=0$ in order to construct a unique configuration
through (\ref{eq:LG2}), with in particular $g_0$ related to the value 
$b_{\rm ext}$ of the external field. From a straightforward linearisation
of the LG equation, one then finds exactly
\begin{equation}
\lim_{f_0,g_0\rightarrow 0}\frac{1}{f_0}f(u)=|L|!\,
\left(\frac{2}{\kappa}\right)^{|L|}\,J_{|L|}(\kappa u)\ \ ,\ \
\lim_{f_0,g_0\rightarrow 0}g(u)=-L\ \ ,\ \ 
\lim_{f_0,g_0\rightarrow 0}b_{\rm ext}=0\ \ \ ,
\label{eq:Bessel}
\end{equation}
where $J_{|L|}(x)$ is the Bessel function of the first kind. In other
words, precisely at the normal-superconducting transition in a vanishing 
external field, there is in fact lurking in the still waters of $f(u)=0$ an
oscillatory pattern just waiting to emerge as soon as $f_0\ne 0$, namely
as soon as the superconducting phase is reached. Thus for $f_0$ arbitrary
close to zero but not vanishing, by continuity there exists a solution
to (\ref{eq:LG2}) and the b.c. (\ref{eq:bcdisk}) at $u=0$ for which
$f(u)$ passes through a succession of zero values while alternating in
sign. In the limit $(f_0=0,g_0=0)$, these values are the zeros of
the Bessel function $J_{|L|}(\kappa u)$. To a good approximation,
their spacing in the sample is such that $\Delta r\approx\pi\xi$, while
the number of those lying within the radius of the sample defines an
upper bound on the maximal number $n_{\rm max}(L)$ of annular vortices possible,
differing from it by at most one unit since the b.c. (\ref{eq:bcdisk})
at $u=u_b$ still need to be solved, and whose value $n_{\rm max}(L)$ thus
decreases as $|L|$ increases.

This remark also leads to the following picture. Having fixed $g_0=0$,
let $f_0$ increase (or decrease, the sign is physically irrelevant)
from $f_0=0$, and generate configurations for $(f(u),g(u))$ using
(\ref{eq:LG2}). Since $f_0$ sets the value for $f(u)$ (when $L=0$)
or for its $|L|$-th order derivative (when $L\ne 0$) at $u=0$, 
and because of the negative
driving force $-\kappa^2f(u)\left[1-f^2(u)\right]$ (for $|f(u)|\le 1$)
in the LG equation, the initial oscillatory pattern
of the Bessel function is then pushed further and further outwards, with the
position of the innermost zero of $f(u)$ moving out while the spacing $\Delta r$
between successive zeros retains more or less the above constant
value (due to the condensate rigidity set by $\xi$). 
Consequently, more and more zeros of $f(u)$ leave the sample until 
configurations with no zeros left within it are reached.
Moreover, throughout the increase of $f_0$ from the null value, there is
a value $f_0^{(n)}$ to be found within each of the successive intervals 
in $f_0$ which are associated to the loss of a zero in $f(u)$
from the sample, such that the b.c. $df(u_b)/du=0$ in (\ref{eq:bcdisk})
is also satisfied. Correspondingly, the solution for $g(u)$ then sets
the associated value for $b_{\rm ext}$ through the second b.c. at $u=u_b$
in (\ref{eq:bcdisk}). Hence in conclusion, with $L$ fixed and $g_0=0$, this
demonstrates the existence of $(L,n)$ annular vortex solutions with
$n=0,1,\cdots,n_{\rm max}(L)$, characterized by having the order parameter
vanish on concentric cylinders when $n\ne 0$
and on the symmetry axis when $L\ne 0$.

In the $(b_{\rm ext},{\cal E}_{\rm norm})$ phase diagram, the above
procedure determines a certain curve starting from the origin\footnote{A series
expansion analysis of the solutions shows that in the case $L=0$,
this curve runs exactly along the negative ${\cal E}_{\rm norm}$ axis,
in which case $g_0=0$ always implies $b_{\rm ext}=0$.}. 
Along that curve, there thus appears in succession a series of points associated 
to the annular vortex solutions with a value of $n$ decreasing from
$n_{\rm max}(L)$ to $n=0$. Each of these points thus also defines a local
extremum of the free energy (\ref{eq:FreeE2}) (which may then take either 
a negative or a positive value), and in fact the numerical
analysis\cite{Gov1} shows that these are local minima in the parameter space
$(f_0,g_0)$ (nevertheless, this is a far cry from having established
dynamical stability for these solutions within the entire configuration
space).

Finally, the complete set of solutions to (\ref{eq:LG2}) with the b.c.
(\ref{eq:bcdisk}) is generated for whatever values of $b_{\rm ext}$
by letting now $g_0$ run up and down its parameter space starting
from $g_0=0$, thereby generating energy curves in the 
$(b_{\rm ext},{\cal E}_{\rm norm})$ phase diagram associated to each of 
the $(L,n)$ annular vortex solutions. This is to be done for each of 
the $(f_0^{(n)},g_0=0)$ solutions
constructed above, by adjusting appropriately the value of $f_0$ for
each new value of $g_0$. To each such solution, there corresponds a specific
value for $b_{\rm ext}$ set by the b.c. in (\ref{eq:bcdisk}). However,
this correspondence between $g_0$ and $b_{\rm ext}$ is not necessarily
one-to-one, as is indeed demonstrated by the numerical analysis. 
Thus, there may exist more than one
solution of fixed $(L,n)$ for some values of the external field, a fact
responsible for hysteresis and switching phenomena between states of different
$L$ values in a varying field,
associated to the crossing points of the negative energy curves of $(L,n)$
vortices in the $(b_{\rm ext},{\cal E}_{\rm norm})$ phase diagram.

The Maxwell equation in (\ref{eq:LG2}) shows that the magnetic field $b(u)$
is stationnary at the zeros of $f(u)$ (there is no Meissner effect
at those points) and that it is monotonous in between zeros of $g(u)$. 
Moreover, since $g(u)\sim uJ(u)/f^2(u)$ is finite at the nodes of $f(u)$,
the supercurrent $J(u)$ must also vanish at these locations while it
remains of constant sign for as long as $g(u)$ does not cross the
null value. In particular, when $L$ and $b_{\rm ext}$ are of opposite sign,
$g(u)$ does not vanish within the sample, so that all the successive
supercurrent vortices retain the same circular orientation.

The same considerations apply in the case of the annular geometry,
the two free b.c. to be adjusted in a likewise fashion
being then the values of $f(u)$ and $g(u)$ at $u=u_a$.
What replaces then the result (\ref{eq:Bessel}) is a linear
combination of the Bessel functions of the first and second kind,
$J_{|L|}(\kappa u)$ and $N_{|L|}(\kappa u)$, while otherwise all the previous
considerations remain applicable, with the obvious provision that the
total number $n$ of zeros in $f(u)$ is reduced as compared to the disk
case because of the inner hole of radius $r_a$.

Hence, the above analytical considerations, though not providing explicit
solutions, demonstrate beyond any doubt the existence of the annular vortex
solutions with the characteristics
described in the Introduction. However, the important issue of their
dynamical stability ({\em i.e.\/} the existence of negative modes
of the quadratic operators in (\ref{eq:LG2})) remains open.
A remark possibly of interest in this respect, to be detailed
elsewhere\cite{Gov1}, is the following. It is well known\cite{Vega}
that for the critical value $\kappa=1/\sqrt{2}$,
the ordinary giant vortex solutions $(L,n=0)$ in the infinite
dimensional plane saturate the Bogomol'nyi bound\cite{Bogo} for the
free energy by satisfying first-order differential equations
from which (\ref{eq:LG2}) then follows. However, given rather the b.c. 
(\ref{eq:bcdisk}) and (\ref{eq:bcann}) associated to a sample of finite
radial extent, it may be shown that these first-order differential
equations are incompatible with these b.c.,
so that the solutions $(L,n\ne 0)$ cannot be con\-si\-de\-red
as being Bogomol'nyi type topological configurations. One may argue that
the solutions $(L,n=0)$ are only a slight deformation of the Bogomol'nyi
configurations implied by the b.c., but this remark does not apply
to the $(L,n\ne 0)$ solutions which are certainly not degenerate in energy
and thus cannot all saturate the same Bogomol'nyi bound. 
Indeed, as soon as $f(u)$ crosses the null value at some radius, from 
there on at larger radii the function $f(u)$ keeps vanishing
at almost regular intervals in an oscillatory pattern, so that the 
free energy of such configurations keeps increasing with increasing
values of $n$ since each successive supercurrent vortex contributes some 
additional amount of kinetic ener\-gy. This is the physical reason 
why annular vortex solutions were never found through formal mathematical 
analyses of the space of solutions to the LG equation in the 
infinite plane\cite{Vega}. 

\noindent{\bf 4. Numerical solutions.}
The above considerations are of course explicitely confirmed through
the numerical resolution of the equations (\ref{eq:LG2}) with the
b.c. (\ref{eq:bcdisk}) and (\ref{eq:bcann}). By lack of space, only
a small---though representative---sample of examples is presented\cite{Gov1}
in the disk case (those for the annulus are very much similar). 
The method followed is the one described above using
the free parameters $(f_0,g_0)$ and a 4$^{\rm th}$ order Runge-Kutta
integration of the differential equations.

Fig.1 displays the $(b_{\rm ext},{\cal E}_{\rm norm})$ phase diagram
for a choice of radius $u_b=r_b/\lambda=5$ and LG parameter
$\kappa=1$. Even though only the curves for $L=0,\pm 1,\pm 2,\pm3$ are shown, 
the $(L,n=0)$ giant vortex configurations follow of course
the usual pattern. Solutions with $n=1$ exist only for $L=0$. Indeed,
for $L\ne 0$, because of the central volume taken up by the magnetic vortex
at the center (see Fig.2), the position of the first zero of $f(u)$ is so 
much pushed outwards that it leaves the sample for this choice of radius. 
For larger values of $\kappa u_b=r_b/\xi$, $n=1$ solutions also exist for 
$L\ne 0$, in which case a series of energy curves similar to those shown in 
Fig.1 for the $n=0$ solutions also appear for the $n=1$ solutions (and so on 
for still larger values of $n$, provided $\kappa u_b$ is large enough; the
choice $\kappa u_b=5$ is made in order that Fig.1 still looks simple enough
to the eye). Clearly, the crossings of all such curves render the dynamics of 
such superconducting devices of large enough radii still far richer and 
more complicated than that provided already by the usual $n=0$ 
configurations\cite{Bruyn2}.

Another feature of Fig.1 worthwhile to emphasize is the caustic-like structure
in the energy curves around their crossing points with the zero energy axis.
The appearance of these cusps stems from the method used to generate
configurations. Solutions move along these curves as the value of $g_0$ varies,
so that the relation between $g_0$ and $b_{\rm ext}$ is not necessarily
one-to-one for some intervals in these quantities. Moreover, solutions
then also move into the positive energy domain, until they meet a return
point back towards the zero energy axis which they reach sometimes after
having moved back into the negative energy domain as well. Quite clearly, this
specific feature displayed by our approach\cite{Gov1} should prove to be
important to an
understanding of the hysteresis phenomena which exist around the critical
fields for the normal-superconducting phase transitions to the $(L,n)$
solutions (even though not apparent on Fig.1, $n\ne 0$ solutions
also possess such caustic behaviour), whose physical properties are determined
by the dynamic and thermodynamic decay rates between the normal and
these different superconducting configurations. These rates are function of
the applied field $B_{\rm ext}$ and the temperature $T$ (note that $\lambda(T)$
and $\xi(T)$ also depend on the latter parameter), as well as the value and
especially the sign of the rate of change in the applied field.

Fig.2 displays all the configurations which exist with $L=0,1$ for a disk 
of radius $u_b=11$ and LG parameter $\kappa=1$, submitted to an external field 
of $b_{\rm ext}=0.05$ (for $\lambda=50$ nm, this corresponds to
$B_{\rm ext}\approx 66$ Gauss). The values of $f(u)$, $b(u)$
and $f^2(u)g(u)/u=q\lambda^3J(u)/\hbar\sim -J(u)$ are shown for 
the $(L=0;n=0,1,2,3)$ and $(L=1;n=0,1,2)$ solutions.
In the latter case, the solution with $n=3$ has disappeared because of the
space taken up by the central magnetic vortex, which thus pushes outwards
the whole train of zeros of the $(L=0,n=3)$ solution. Note also that
for this disk with $u_b=11$ and $\kappa=1$, 
the values of each of the solutions $n=0,1,2$ coincide almost exactly 
beyond $u>0.4 u_b$ when comparing the $L=0$ and $L=1$ configurations, since 
the effects of the $L=1$ magnetic vortex remain confined within a radius of 
a few penetration lengths $\lambda$. Further properties of these configurations
have been discussed in general terms already, in particular the
fact that the magnetic field $b(u)$ is stationnary at nodes of the order
parameter $\psi(r,\phi)$, where the supercurrent $J(u)$ then also vanishes, 
thereby enabling further penetration of the external field within the
superconducting sample and a partial anti-screening of the Meissner effect
for $n\ne 0$ annular vortices as compared to the usual $n=0$ configurations.
These few examples thus confirm exactly the results advocated on basis
of the general analytical considerations developed previously.

\noindent{\bf 5. Conclusions.}
This letter has demonstrated the existence of annular vortex solutions
to the Landau-Ginzburg equations in mesoscopic superconductors. These
configurations include and generalise the well known Abrikosov
and giant magnetic vortex ones, are characterized by properties des\-cribed 
above, and exist only thanks to the finite spatial extent of mesoscopic devices. 
Important issues remain open obviously, most of them raised
by the eventual experimental observation of direct or indirect physical 
effects related to the existence in actual samples of such annular vortices,
either as single entities or organised into regular or irregular ensembles.
Dynamic and thermodynamic stability is obviously essential for their
direct observation, and remains to be investigated. Irrespective of their
stability, these configurations should have indirect manifestations since
they enrich the phase diagram of mesoscopic superconducting
disks and loops in an external magnetic field in ways still to be
explored. Besides hysteresis phenomena in such systems to which our
approach should be applicable, annular
vortices also provide for a new mechanism for the switching 
between states of different fluxoid quantum number, which---since these solutions
manifestly share the cylindrical symmetry of the sample---is in direct
competition with the specific me\-cha\-nism unravelled in Ref.\cite{Berger}. 
Static and dynamic properties of mesoscopic superconducting
devices should thus be affected by the existence of such annular vortices,
in ways to be assessed explicitely and whose conclusions could
potentially be of importance when having specific practical applications 
of such devices in mind. We plan to report on progress on some of these 
issues in later work.

\noindent{\bf Acknowledgments.}
Profs. Vincent Bayot, Christian Fabry and Ghislain Gr\'egoire are 
gratefully acknowledged for their interest in this work, and especially
the latter two for their constructive suggestions towards the numerical 
resolutions of the equations. In particular, Chr.~Fabry showed us how
to efficiently apply Mathematica tools to our problem.
This work is part of the Master's Diploma Thesis of G.S. and the
undergraduate Diploma Theses of D.B. and O.vdA.
The work of G.S. is financially supported 
as a Scientific Collaborator of the
``Fonds National de la Recherche Scientifique" (FNRS, Belgium).

\clearpage

\noindent{\bf Figure Captions}

\vspace{10pt}

\noindent Figure 1: The normalised free energy ${\cal E}_{\rm norm}$
of (\ref{eq:FreeE2}) as a function of the normalised applied field
$b_{\rm ext}$ for a disk with $u_b=5$ and $\kappa=1$. Displayed are
the $n=0$ giant vortex configurations with $L=-3,-2,-1,0,1,2,3$ from left to
right in that order (thin parabolic curves), while the $n=1$ solution 
exists for $L=0$ only (thick line).

\vspace{10pt}

\noindent Figure 2: The $(L=0;n=0,1,2,3)$ and $(L=1;n=0,1,2)$ configurations
for a disk with $u_b=11$, $\kappa=1$ and $b_{\rm ext}=0.05$. Displayed
in that order from left to right
are the values for $f(u)$, $b(u)$ and 
$f^2(u)g(u)/u=q\lambda^3J(u)/\hbar\sim -J(u)$ as functions of the
variable $x=u/u_b=r/r_b$, $0\le x\le 1$. The top (resp. bottom) panels
correspond to $L=0$ (resp. $L=1$). The $f(u)$ values for the $(L=0,n=0)$ 
solution coincide almost exactly with $f(u)=1$ and cannot be distinguished 
from unity on these graphs (one has $f(u_b)=0.9995072$ and $f(u_b)=0.9995076$ 
for $(L=0,n=0)$ and $(L=1,n=0)$, respectively).

%\begin{figure}[ht]
%\begin{center}
%\vspace*{3mm}
%\includegraphics[height=12.5cm]{Fig1.eps}
%\caption{The normalised free energy ${\cal E}_{\rm norm}$
%of (\ref{eq:FreeE2}) as a function of the normalised applied field
%$b_{\rm ext}$ for a disk with $u_b=5$ and $\kappa=1$. Displayed are
%the $n=0$ giant vortex configurations with $L=-3,-2,-1,0,1,2,3$ from left to
%right in that order (thin parabolic curves), while the $n=1$ solution 
%exists for $L=0$ only (thick line).}
%\vspace*{-3mm}
%\end{center}
%\end{figure}
%
%\clearpage
%
%\begin{figure}[ht]
%%\begin{center}
%\vspace*{3mm}
%\includegraphics[width=17cm]{Fig2.eps}
%%\includegraphics[width=17cm,height=11.5cm]{Fig2.eps}
%\caption{The $(L=0;n=0,1,2,3)$ and $(L=1;n=0,1,2)$ configurations
%for a disk with $u_b=11$, $\kappa=1$ and $b_{\rm ext}=0.05$. Displayed
%in that order from left to right
%are the values for $f(u)$, $b(u)$ and 
%$f^2(u)g(u)/u=q\lambda^3J(u)/\hbar\sim -J(u)$ as functions of the
%variable $x=u/u_b=r/r_b$, $0\le x\le 1$. The top (resp. bottom) panels
%correspond to $L=0$ (resp. $L=1$). The $f(u)$ values for the $(L=0,n=0)$ 
%solution coincide almost exactly with $f(u)=1$ and cannot be distinguished 
%from unity on these graphs (one has $f(u_b)=0.9995072$ and $f(u_b)=0.9995076$ 
%for $(L=0,n=0)$ and $(L=1,n=0)$, respectively).}
%\vspace*{-3mm}
%%\end{center}
%\end{figure}


\clearpage

\newpage

\begin{thebibliography}{99}

\bibitem{Tin} For reviews and references to the original literature,
see for example,\\
M. Tinkham, {\sl Introduction to Superconductivity\/},
2nd edition (McGraw-Hill, New York, 1996);\\
J.R. Waldram, {\sl Superconductivity of Metals and Cuprates\/}
(Institute of Physics, Bristol, 1996).

\bibitem{Tris} T. Alexander, Yu.S. Kivshar, A.V. Buryak and R. Sammut,
{\sl Optical Vortex Solutions in Parametric Wave Mixing\/},
preprint {\tt patt-sol/9904004} (April 1999).

\bibitem{Vega} H.J. de Vega and F.A. Schaposnik, {\sl Phys. Rev.\/} {\bf 14}
(1976) 1100;\\
L. Jacobs and C. Rebbi, {\sl Phys. Rev.\/} {\bf B19} (1979) 4486;\\
E. Weinberg, {\sl Phys. Rev.\/} {\bf D19} (1979) 3008;\\
C.H. Taubes, {\sl Comm. Math. Phys.\/} {\bf 72} (1980) 277; 
{\sl ibid} {\bf 75} (1980) 75;\\
Yi. Yang, {\sl Comm. Math. Phys.\/} {\bf 123} (1989) 123.

\bibitem{Gustafson} S. Gustafson and I.M. Sigal, {\sl The stability of
magnetic vortices\/}, preprint {\tt math.AP/9904158} (April 1999).

\bibitem{Geim1} A.K. Geim, S.V. Dubonos, J.G.S. Lok, M. Henini and
J.C. Maan, {\sl Nature\/} {\bf 396} (1998) 144.

\bibitem{Bruyn1} Xi. Zhang and J.C. Price, {\sl Phys. Rev.\/} {\bf B55}
(1997) 3128;\\
V.V. Moshchalkov, X.G. Qiu and V. Bruyndoncx,
{\sl Phys. Rev.\/} {\bf B55} (1997) 11793.

\bibitem{Bruyn2} For a representative list of references, see\\
V. Bruyndoncx, L. Van Look, M. Verschuere and 
V.V. Moshchalkov, {\sl Dimensional crossover in a mesoscopic superconducting
loop of finite width\/}, preprint {\tt cond-matt/9907341} (July 1999);\\
V. Bruyndoncx {\em et al.\/}, {\sl Giant vortex state in
perforated aluminium microsquares\/}, preprint {\tt cond-matt/9905093}
(May 1999).

\bibitem{Gov1} J. Govaerts, G. Stenuit, D. Bertrand and O. van der Aa,
in preparation.

\bibitem{Bogo} E.B. Bogomol'nyi, {\sl Sov. J. Nucl. Phys.\/}
{\bf 23} (1976) 588.

\bibitem{Berger} J. Berger and J. Rubinstein, 
{\sl Phys. Rev.\/} {\bf B59} (1999) 8896.

\end{thebibliography}
\end{document}